\pdfoutput=1

\documentclass[draft,jgrga]{agutex2015}

 \usepackage{graphicx}
 \usepackage{bmpsize}
  \setkeys{Gin}{draft=false}
\authorrunninghead{AMERSTORFER ET AL.}

\titlerunninghead{Evolution of Mars' CO$_2$ atmosphere}

\begin{document}

%
%

\title{Escape and evolution of Mars' CO$_2$ atmosphere: Influence of suprathermal atoms}
%
%

%
%



\authors{U. V. Amerstorfer,\altaffilmark{1}
 H. Gr\"oller,\altaffilmark{2} H. Lichtenegger,\altaffilmark{1}
H. Lammer,\altaffilmark{1} F. Tian,\altaffilmark{3} L. Noack,\altaffilmark{4} M. Scherf,\altaffilmark{1} C. Johnstone,\altaffilmark{5} L. Tu,\altaffilmark{5} and M. G\"udel\altaffilmark{5}}

\altaffiltext{1}{Space Research Institute, Austrian Academy of Sciences, Graz, Austria.}

\altaffiltext{2}{Lunar and Planetary Laboratory, University of Arizona, Tucson, AZ, USA.}

\altaffiltext{3}{Ministry of Education Key Laboratory for Earth System Modeling, Center for Earth System Science, Tsinghua University, Beijing, China.}

\altaffiltext{4}{Royal Observatory of Belgium, Brussels, Belgium.}

\altaffiltext{5}{Department of Astrophysics, University of Vienna, Vienna, Austria.}

%
%


\keypoints{
\item Escape of suprathermal atoms from Mars' atmosphere with increasing EUV flux is studied
\item Mars could not have had a dense atmosphere at the end of the Noachian epoch
\item Mars' surface pressure could have been larger than 1 bar during the first 300 Myr after the planet's origin}


%
%


\begin{abstract}
With a Monte-Carlo model we investigate the escape of hot oxygen and carbon from the martian atmosphere for four points in time in its history corresponding to 1, 3, 10, and 20 times the present solar EUV flux. We study and discuss different sources of hot oxygen and carbon atoms in the thermosphere and their changing importance with the EUV flux. The increase of the production rates due to higher densities resulting from the higher EUV flux competes against the expansion of the thermosphere and corresponding increase in collisions. We find that the escape due to photodissociation increases with increasing EUV level. However, for the escape via some other reactions, e.g.~dissociative recombination of O$_2^+$, this is only true until the EUV level reaches 10 times the present EUV flux, and then the rates start to decrease. Furthermore, our results show that Mars could not have had a dense atmosphere at the end of the Noachian epoch, since such an atmosphere would not have been able to escape until today. In the pre-Noachian era, most of a magma ocean and volcanic activity related outgassed CO$_2$ atmosphere could have been lost thermally until the Noachian epoch, when non-thermal loss processes such as suprathermal atom escape became dominant. Thus, early Mars could have been hot and wet during the pre-Noachian era with surface CO$_2$ pressures larger than 1 bar during the first 300 Myr after the planet's origin.
\end{abstract}

%
%

%

\begin{article}

\section{Introduction}

Throughout their evolution, planetary atmospheres are strongly influenced by the radiation and particle emissions from their host star. Different studies have shown that the Sun's radiation in the extreme ultraviolet (EUV) part of the solar spectrum was higher in the past \citep{ribas05,tu15}, and thus, the planetary atmospheres are exposed to varying external conditions. \citet{tu15} have shown that the star's initial rotation rate and its rotational evolution play an important role for the EUV flux enhancement and the evolution of the atmospheres of terrestrial planets.

The time scales of different epochs in the martian history depend on the evolution of the solar EUV flux \citep[e.g.][]{tu15}. As shown by \citet{tian09} for EUV fluxes higher than about 20 times that of today's Sun, Mars' CO$_2$ atmosphere will experience high thermal loss rates, such that one cannot expect the buildup of a dense atmosphere \citep{lammer13,erkaev14}. Because of the high EUV flux ($\ge$ 20 EUV) of the young Sun, a large fraction of IR-cooling molecules in the thermosphere, such as CO$_2$ (also a greenhouse gas), has been dissociated. Due to Mars' low gravity, the upper atmosphere expanded hydrodynamically, so that hydrodynamic blow-off of hydrogen and strong thermal escape rates of heavier species, such as O and C atoms, occurred independently of the early Mars' magnetic field \citep{tian09}.

There is geomorphologic evidence \citep[e.g.][]{craddock93,malin03} that the early martian climate during the Noachian ($4.1 - 3.7$ Gyr ago) and Hesperian ($3.7 - 2.0$ Gyr ago) epochs should have been warm enough for liquid water flow, at least sporadically, on the surface. Remote sensing from Mars orbiters and in situ analyses of the surface mineralogy by Mars rovers indicate also the local presence of minerals such as clay/phyllosilicates, clathrates, opalia silica, sulphates, and chlorides, which require liquid water for their formation \citep[e.g.][]{gendrin05,bibring06,osterloo08,squyres08}. From these observations it is clear that the environmental conditions on early Mars varied substantially throughout the Noachian and Hesperian era.

However, it is unknown if the conditions suitable for liquid water were stable on longer timescales, or if they were the consequences of episodic, possibly catastrophic events.  Recent sophisticated 3D global climate simulations of the early martian atmosphere suggest that the annual mean temperature could not have reached values above 0$^{\circ}$ C anywhere on the planet by a CO$_2$ atmosphere and that the conditions do not allow long-term liquid water on the surface \citep{forget13,wordsworth13}. The models predict a collapse of the atmosphere into permanent CO$_2$ ice caps for pressures higher than 3 bar, or, if the obliquity is low enough, for pressure values less than 1.0 bar. These model results indicate a cold early Mars, where nonclimate processes have to be responsible for liquid water. Other studies included greenhouse effects by gases such as CH$_4$ and SO$_2$ \citep{johnson08,halevy14}. These gases are also unable to rise the surface temperature because CH$_4$ causes stratospheric warming at the expense of surface warming \citep{ramirez14} and SO$_2$ forms sulfate aerosols which act as coolers for the climate, too \citep{tian10,kerber15}. The latest hypothesis on the martian climate is related to global cirrus cloud decks in a CO$_2$-H$_2$O atmosphere with at least $250$ mbar of CO$_2$, which might have been able to keep Mars warm \citep{urata13}. Recently, \citet{ramirez16} showed that this process works only for special cloud properties and if cirrus clouds cover about $75 - 100$ \% of the planet. Therefore, these authors conclude that it is most likely that the cirrus cloud hypothesis does not provide the necessary warming, which indicates the need for other greenhouse mechanisms.

In the early pre-Noachian epoch after the solidification of an early martian magma ocean, as it was shown by \citet{erkaev14}, catastrophically outgassed volatiles with the amount of $50 - 250$ bar H$_2$O and about $10 - 55$ bar CO$_2$ \citep{elkins-tanton08} could have been lost during the EUV saturation period of the young Sun, if the EUV flux was larger than 100 times the present solar value. Especially, if the majority of CO$_2$ molecules had been dissociated and if the impact related energy flux of large planetesimals and small embryos to the planet's surface lasted long enough, the steam atmosphere could have been prevented from condensing \citep{maindl15}. However, if the solar EUV flux was lower, so that CO$_2$ molecules might not have been effectively dissociated or dragged away by the thermally escaping hydrogen flux, then the results suggest  temporary periods (e.g. through impacts or episodic volcanic outgassing), where some amount of liquid water might have been present on the planet's surface.

Besides the catastrophic outgassing due to magma ocean solidification, later degassing by volcanic processes could have built up a secondary CO$_2$ atmosphere during the Noachian and Hesperian epochs \citep[e.g.][]{phillips01,grott11,lammer13}. Investigating such a scenario, \citet{tian09} assumed volcanic outgassing rates in analogy to Earth based on studies of \citet{phillips01}. In such cases, high atmospheric CO$_2$ surface pressures of the order of about 1.5 bar are obtained for the outgassing associated with the formation of Tharsis alone \citep{phillips01}. Assumptions of higher volatile amounts as in \citet{tian09}, which yield atmospheres with a surface pressure of about 3 bar during the late Noachian/early Hesperian era, do not agree with recent paleopressure studies of \citet{kite14}. These authors considered the estimates from the size-frequency distribution of small ancient craters ($>$ 3.6 Gyr ago) interspersed with river deposits in the Aeolis region in combination with simulations of the effect of atmospheric pressure on the crater flux and obtained upper limits for the surface pressure values of approximately $0.9 - 1.9$ bar (depending on model assumptions). In agreement with \citet{kite14}, \citet{hu15} found that the current atmospheric isotope ratio $^{13}$C/$^{12}$C and carbonate measurements hint at an early martian atmosphere with less than 1 bar surface pressure. They considered sputtering and photochemical escape of C and extrapolated the current escape rates to the past. They underline, however, that the understanding of how the photochemical escape rates change with the EUV flux is of great importance to minimize the uncertainties in the surface pressure estimates.

Moreover, \citet{hirschmann08} and \citet{stanley11} reinvestigated the CO$_2$ content of martian magmas and found that, apart from pressure and temperature conditions in the magma source regions, the CO$_2$ content primarily depends on oxygen fugacity. For a plausible range of pressure-temperature conditions, their estimates yield CO$_2$ contents in martian magmas of about $0.01 - 0.1$ wt\%, indicating that the magmatic CO$_2$ content has probably been overestimated in previous models. A lower CO$_2$ content reduces of course the amount of CO$_2$ released to the atmosphere by volcanoes. \citet{hirschmann08} estimated that approximately $0.1 - 1$ bar of CO$_2$ could have been outgassed after 0.7 Gyr, which agrees well with the lower value of the paleopressure estimates by \citet{kite14}.

Based on the findings of \citet{hirschmann08}, \citet{grott11} applied a parameterized thermo-chemical evolution model \citep{morschhauser11} for volcanic outgassing on early Mars by considering two end-member melting models. They calculated self-consistently the amount of outgassed CO$_2$ and H$_2$O during the martian evolution. It was found that outgassing rates depend primarily on the bulk mantle water content, the mantle oxygen fugacity, and the local melt fraction in the magma source regions. By assuming a global melt channel, outgassing is most efficient in the pre-Noachian era (up to 4.1 Gyr ago: about 750 mbar), but still significant during the Noachian epoch, where about 250 mbar of CO$_2$ could be outgassed between 4.1 and 3.7 Gyr ago. Outgassing decreases significantly during the Hesperian ($3.7 - 2$ Gyr ago: about 20 mbar) and Amazonian (2 Gyr ago $-$ present) eras. In case one assumes that melting is restricted to localized mantle plumes, then approximately 240 and 365 mbar could have been outgassed during the Noachian and Hesperian epochs, respectively. For both melt channel scenarios a total of approximately $0.9 - 1$ bar CO$_2$ could have been outgassed by volcanoes.

From these times until today, the martian atmosphere was most likely modified by a complex interplay of escape by suptrathermal atoms \citep[e.g.][]{fox04}, sputtering \citep{jakosky94}, ion escape, impacts, carbonate precipitation, and serpentinization \citep{chassefiere11}, which led to the present-day surface pressure. \citet{zhao15} calculated photochemical escape of oxygen from a Mars atmosphere exposed to 1, 3, 10, and 20 times the present solar EUV flux. They focussed on dissociative recombination of O$_2^+$ as source of the energetic oxygen atoms. \citet{groeller14}, however, have shown that dissociative recombination of CO$_2^+$ also plays an important role in producing hot oxygen atoms in the present martian atmosphere. This finding was recently confirmed by \citet{lee15}. In the present work, we extend the study of \citet{zhao15} by including more source reactions for hot oxygen and by including hot carbon, since both species are connected to the loss of CO$_2$ throughout the martian evolution. We also discuss thermal loss in the very early times of Mars and several volcanic outgassing scenarios, which represent a CO$_2$ input to the atmosphere. \citet{gillmann09} and \citet{gillmann11} similarly studied the combined effect of volcanic outgassing and atmospheric escape on the atmosphere of Mars. They extrapolated loss rates of different processes to the past and used the scaling laws by \citet{ribas05} to relate different EUV levels to times in the past. The results of \citet{gillmann11} suggest that the martian atmospheric pressure was around 50 mbar 4 Gyr ago and that today's atmosphere consists to a large part of volcanic gases.

With this study we want to address the following questions:
\begin{itemize}
  \item How does the importance of different production reactions of hot O and hot C vary with a higher EUV flux?
  \item How much CO$_2$ can approximately be lost through suprathermal atoms since the Noachian era?
  \item How does loss through suprathermal atoms influence the evolution of the martian surface pressure?
\end{itemize}
In Section \ref{sec:model} we describe the Monte-Carlo model, the atmospheric input parameters and the production reactions used to study the escape of hot O and C from the atmosphere of Mars for different EUV flux exposures. Section \ref{sec:results} presents the results of our investigation, and Section \ref{sec:surfPress} discusses implications for the surface pressure evolution. A summary and conclusions of our findings are presented in Section \ref{sec:summary}.

\section{Model description and input} \label{sec:model}

\subsection{Monte-Carlo model}

Energetic hot atoms in the atmosphere of Mars are produced via different source reactions. For a specific reaction, we determine the corresponding velocity distribution for the reaction products at discrete altitudes. We follow these products along their 3-dimensional path through the thermosphere up to the exobase in the gravitational field of Mars. On their way, the hot particles interact with the background neutral atmosphere via collisions and lose on average part of their initial energy, whereas background particles gain energy through such collisions and may thus become hot. We adopt the following strategy for considering new secondary hot particles: Denoting the energy of the background particle before and after the collision by $E$ and $E'$, respectively, a new secondary hot particle is created if $E<1.5\,E_{\rm{therm}}$ and $E'>1.5\, E_{\rm{therm}}$ (with $E_{\rm{therm}}$ being the thermal energy). When $E>1.5\, E_{\rm{therm}}$, the particle is only considered as hot if $E'>1.5\, E$. The factor $1.5$ is a compromise between taking the relevant particles with high enough energies and simulation time. However, the factor is chosen such that the error, when excluding some newly produced secondary hot particles, is not significant, since those particles do not have high enough energies to be able to escape. In fact, for EUV levels of 1 to 3 times the present solar EUV flux, we could even increase the factor and we would still be an order of magnitude below the relevant escape energy at the exobase.

The collision probability and the energy transfer is calculated by means of total and differential cross sections. At the exobase altitude, the energy distribution function of the suprathermal particles is determined, which in turn serves as input for the exosphere density and loss rate calculations. More details of the Monte-Carlo model can be found in \citet{groeller10,groeller12,groeller14}.

\subsection{Input profiles}\label{sec:input}

\begin{figure}
 \noindent\includegraphics[width=0.9\textwidth]{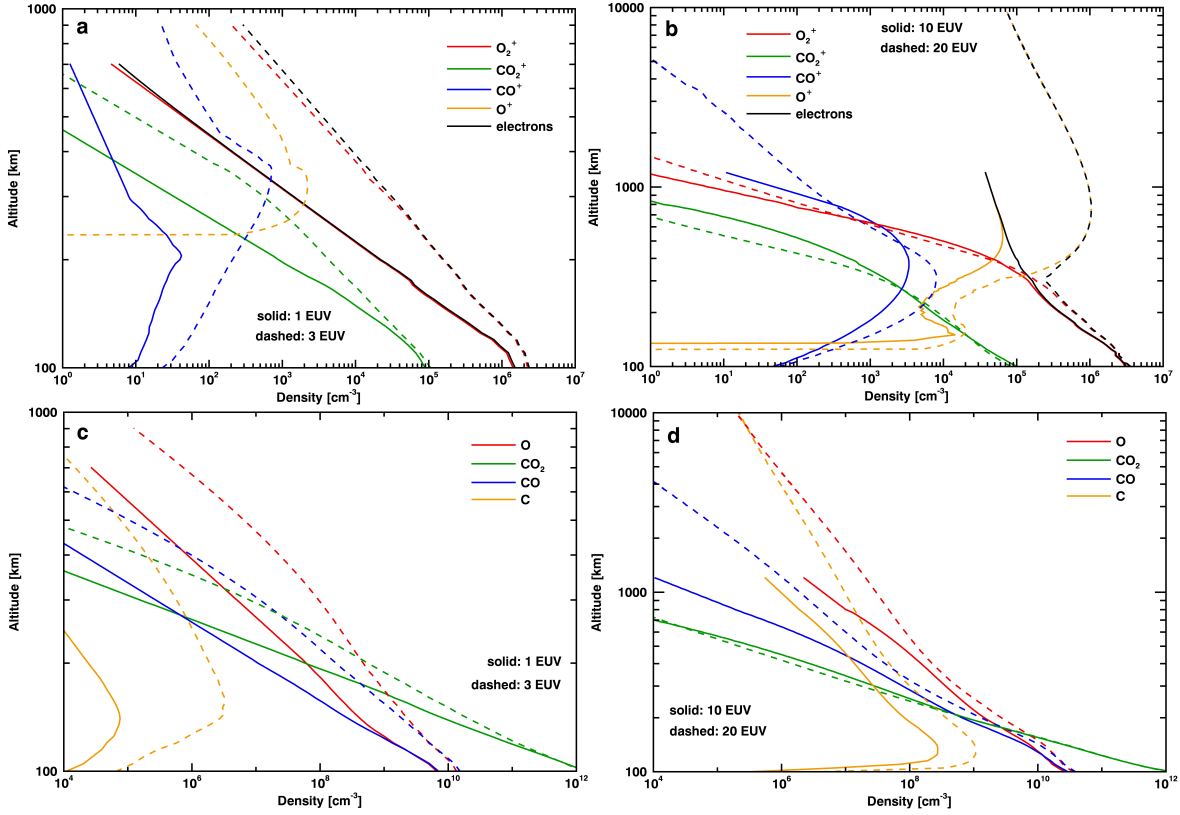}
 \caption{Ion, electron and neutral profiles for the four EUV flux cases. Panel (a) shows the ion densities for the 1 (solid) and 3 EUV (dashed) cases. Please note that there is no O$^+$ density for the 1 EUV case. Panel (b) displays the ion densities for the 10 (solid) and 20 EUV (dashed) cases. In panel (c), the neutral densities for the 1 (solid) and 3 EUV (dashed) cases are shown, whereas panel (d) shows the neutral densities for 10 (solid) and 20 EUV (dashed) (adopted from \citet{tian09}).}
 \label{fig:Prof}
 \end{figure}

\begin{figure}
 \noindent\includegraphics[width=0.9\textwidth]{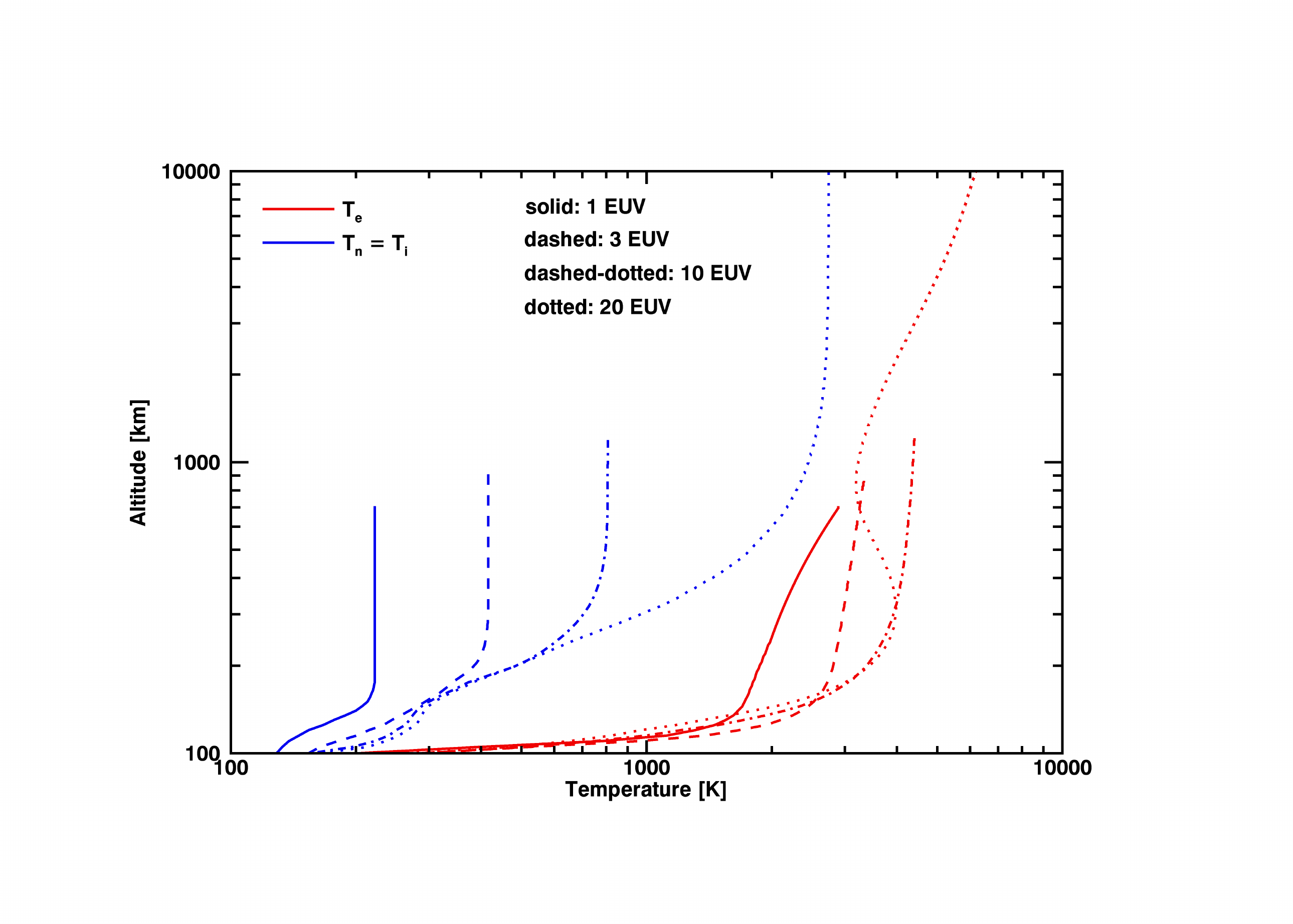}
 \caption{ Temperature profiles for the four EUV flux cases. The ion temperature is assumed to be equal to the temperature of the neutral species (adopted from \citet{tian09}).}
 \label{fig:TempProf}
 \end{figure}

Figure \ref{fig:Prof} shows the atmospheric profiles of O, CO, CO$_2$, C, O$_2^+$, CO$_2^+$, CO$^+$, and O$^+$ used as background atmosphere for our Monte-Carlo simulations. These profiles were adopted from \citet{tian09}, who simulated the martian atmosphere for four different solar EUV fluxes, namely 1, 3, 10, and 20 times the today's flux (henceforth, we will denote these four cases as 1, 3, 10, and 20 EUV cases). In their calculations, the solar EUV radiation is divided by a factor of 4 to account for global mean profiles. Differences of the profiles in the ionosphere below 150 km compared to other studies \citep[e.g.][]{fox09} arise mainly from the lower boundary conditions. However, since hot atoms produced in such low altitudes do not have a high escape probability and since our Monte-Carlo simulation starts at 150 km, these differences have no significant influence on our results. The lower boundary condition for the ionosphere is even less important for the ionospheric structure for elevated EUV conditions. The enhancement of the EUV flux leads to an increase in exobase temperature and altitude, which becomes stronlgy significant for the 20 EUV case. The dip in the electron density for the 20 EUV case (Figure \ref{fig:Prof}) results from the transition of an O$_2^+$ dominated ionosphere to an O$+$ dominated one. Figure \ref{fig:TempProf} shows the temperature profiles for the four EUV cases. The temperature of the ions is assumed to be the same as the temperature of the neutral species. The dip in the electron temperature for the 20 EUV case is due to a strong coupling of the neutrals, ions and electrons in these altitudes. The model, with which the background profiles were obtained, is based upon the Earth model presented in \citet{tian08a} and was used to simulate the martian atmosphere in \citet{tian09} and \citet{zhao15}, in which a subset of the shown profiles is presented. Neither \citet{tian09} nor \citet{zhao15} provide an O$^+$ profile for the 1 EUV case. However, for our study, O$^+$ does not play a role in producing hot O atoms, since we do not inlcude any reaction involving O$^+$. From observations it is also clear that O$^+$ does not contribute significantly to the electron density in the altitudes that we are interested in, i.e.~below 220 km \citep[e.g.][]{benna15}. Thus, the neglection of O$^+$ in the 1 EUV case does not alter our results.

\subsection{EUV flux}
\begin{figure}
 \noindent\includegraphics[width=0.9\textwidth]{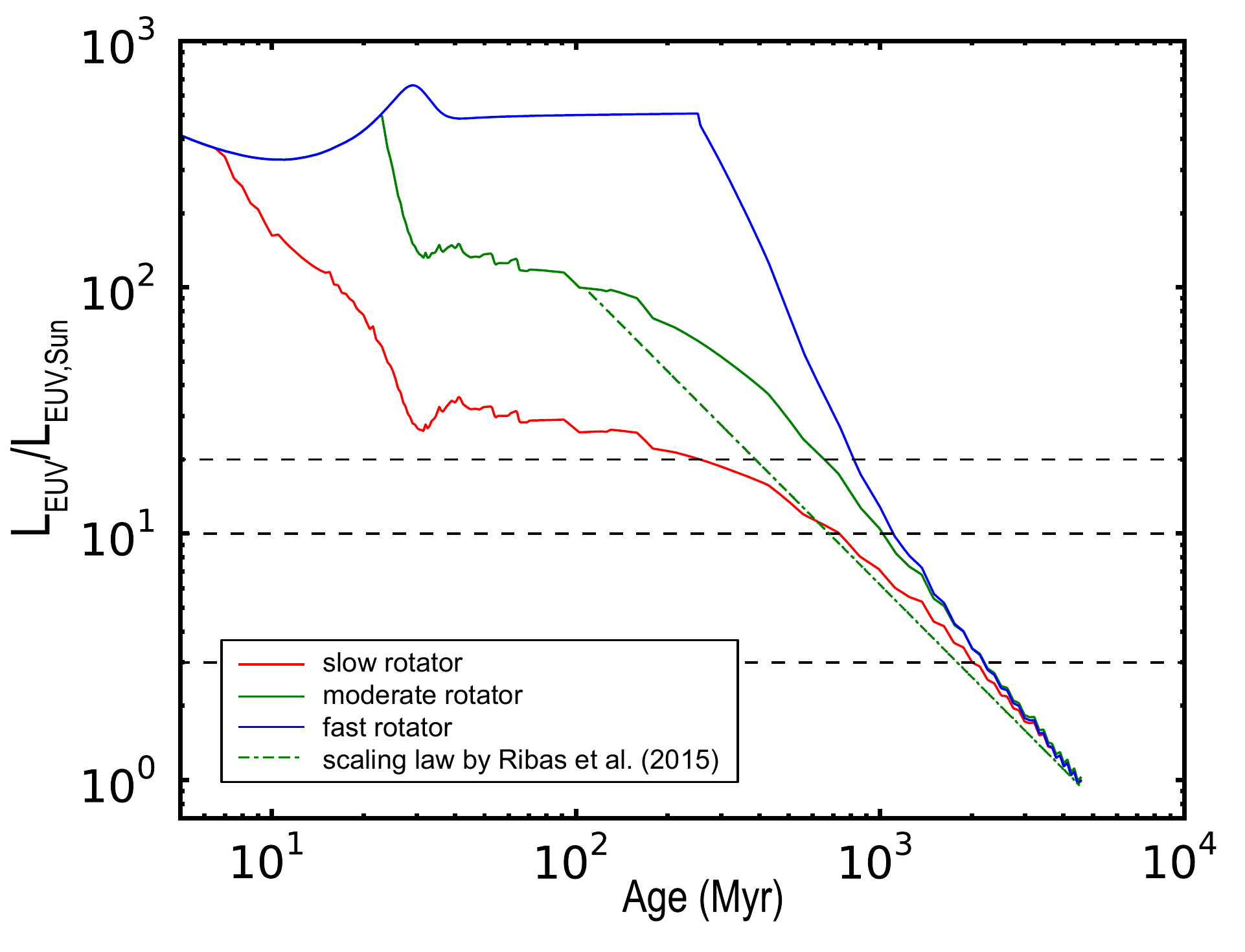}
 \caption{Radiation evolution of three different rotator types of stars. For comparison, the scaling law by \citet{ribas05} is also shown. L$_{\rm{EUV}}$ is the EUV flux of the star and L$_{\rm{EUV,Sun}}$ is the present EUV flux of the Sun. The horizontal black dashed lines indicate three different EUV levels: 3, 10, and 20 times the present solar EUV flux (adapted from \citet{tu15}).}
 \label{fig:radEvol}
 \end{figure}

To what times in the past do such solar EUV levels correspond? As pointed out by \citet{johnstone15b}, it is quite common to use the scaling laws of \citet{ribas05} to estimate the evolution of the Sun's radiation in EUV. However, the radiation of a star depends, apart from other things, on its rotational evolution \citep{johnstone15b,tu15}. About 70\% of the solar mass stars examined by \citet{johnstone15b} are slow rotators, whereas there is a non-negligible possibility for the Sun to have been a fast rotator, giving a completely different rotational, and thus radiation, evolution. The EUV flux evolution model of \citet{tu15} takes into account a broad observational sample of stars in clusters with ages from 30 Myr to 620 Myr. High energy radiation of a star decreases over time as a result of rotational spin-down. Due to the high amount of observed stars, the Tu-model can set the EUV evolution in correlation to the initial rotation rate of a star, in contrast to the Ribas-model due to its limited sample of stars. Since the Tu-model covers much wider evolution scenarios for stars, it is used in the present work for estimating the rotational evolution of the Sun. We thus consider three cases, slow, moderate, and fast rotators, to determine the times in the past, corresponding to each EUV level case, as shown in Table \ref{tab:rotEUVTimes}. Figure \ref{fig:radEvol} illustrates the radiation evolution of these three rotator types obtained with the model of \citet{tu15}, together with the law of \citet{ribas05}. The radiation evolution model of \citet{tu15} is an extension of the rotational evolution model by \citet{johnstone15b} and is strongly dependend on the initial rotation rate of the star. To constrain their rotational model, they assume that the percentiles of the rotational distributions for star clusters of different ages can be combined to derive the rotational evolution of a solar-like star. The rotational evolution model always refers to the equatorial plane and does not take averaged values for the rotation. Thus, rotation period, solar wind velocity and density, as well as solar magnetic field retrieved with this model are only valid for the equatorial plane. To predict the EUV flux along the different rotational tracks, \citet{tu15} use the relation between Rossby-number and X-ray luminosity as derived by \citet{wright11}, and the power law by \citet{sanzforcada11} for converting the X-ray luminosity into the EUV flux.

\subsection{Source reactions and production rates}

Dissociative recombination (DR) of O$_2^+$ and CO$_2^+$ are the main sources of hot oxygen in the atmosphere of present Mars \citep[e.g.][]{fox09,groeller14,lee15}. In \citet{groeller14} the escape of hot oxygen and carbon from the present Mars atmosphere was studied. Table \ref{tab:sources} summarizes the production reactions and their corresponding rate coefficients for the production of hot O and hot C considered in this study. In addition, we also include photodissociation (PD) of CO as sources for hot oxygen and hot carbon. Although some of these reactions are of little importance for present day Mars, they turn out to be important in earlier times due to the larger EUV flux.

For dissociative recombination, the rate coefficient reads

\begin{equation}
k (T_{\rm{e}}) = \alpha\, \left(\frac{T_{\rm{e}}}{300}\right)^{\beta}\, \textrm{cm}^3 \,\textrm{s}^{-1},
\end{equation}

where $T_{\rm{e}}$ is the electron temperature in Kelvin. For the chemical reaction O$_2^+$ + C $\rightarrow$ CO$^+$ + O, the rate coefficient is assumed to be constant and does not depend on the neutral temperature. The values used for $\alpha$ and $\beta$ are listed in Table \ref{tab:sources}. As in \citet{groeller14} we took a branching ratio of $4\%$ for the reaction CO$_2^+$ + e $\rightarrow$ O$_2$ + C, which is the maximum branching ratio as given by \citet{viggiano05} and gives the maximum contribution for the production of hot C from this source reaction.

For photodissociation, the solar flux is taken from SUMER/SOHO observations \citep{curdt01,curdt04}. We have chosen observations from April 20, 1997, for quiet Sun conditions. Details about the instrument, its detectors, the observed solar spectrum, its calibration and the spectrum itself can be found in \citet{curdt01} and \citet{curdt04}. We converted the data to the units of photons cm$^{-2}$ s$^{-1}$ A$^{-1}$ and transferred them to the orbit of Mars by dividing the photon flux by the square of the Sun-Mars distance in AU. The Chapman function for an isothermal atmosphere is used to adopt the solar flux to the considered solar zenith angle. We assume the input profiles to represent average dayside conditions and thus the solar zenith angle is taken to be 60$^\circ$. The altitude dependent production rate $P(r)$ for photodissociation of a neutral molecule is calculated by

\begin{equation}
 P(r) = n_{\rm{s}}(r) \int_\lambda \sigma_{\rm{s}}^{\rm PD}(\lambda) F(r,\lambda) d\lambda \,,
\end{equation}

where $n_{\rm{s}}(r)$ is the altitude dependent density of the neutral species, $F(r,\lambda)$ the solar flux for a given wavelength $\lambda$ at the altitude $r$, and $\sigma_{\rm{s}}^{\rm PD}(\lambda)$ the photodissociation cross section for the neutral at wavelength $\lambda$. The photodissociation and absorption cross sections are taken from \citet{huebner92}, who provide the data in the Photo Rate Coefficient Database.

\begin{figure}
 \noindent\includegraphics[width=0.9\textwidth]{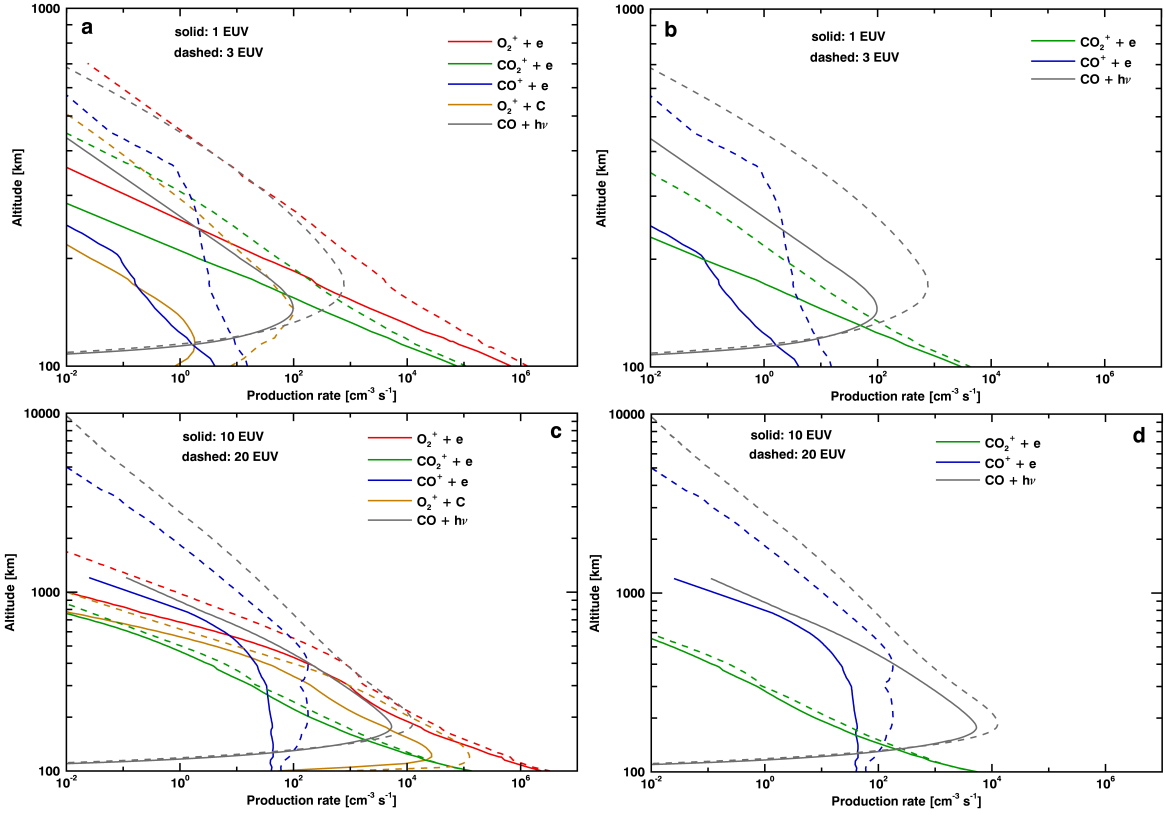}
 \caption{Production rates for hot oxygen (panel a and c) and hot carbon (panel b and d) for different reactions and different EUV fluxes.}
 \label{fig:prod}
 \end{figure}

The production rates of hot O and hot C are shown in Figure \ref{fig:prod}. Due to the lower boundary conditions of the atmosphere model, some reactions do not show a maximum in their production rate profiles. However, the values at 150 km, where we start our simulation, and above are comparable to previous studies \citep[e.g.][]{groeller14}. As shown and discussed in \citet{groeller14}, it is not for all reactions true that the production rates are higher for high solar activity than for low solar activity. Basically, the variation of the input profiles (neutral and ion profiles) determines the variation of the production rates. The input profiles used in this study for the 1 EUV case lie in between high and low solar acitivy profiles of other studies \citep[e.g.][]{fox09}.

 The kinetic energy attained by the hot atom via a DR source reaction is randomly chosen from the energy distribution of this reaction. The energy distribution is obtained from the total kinetic energy in the center of mass frame, which is given by E$_{\rm{tot}}$ = E$_{\rm{cm}}$ + E$_{\rm{br}}$ + E$_v$ + E$_r$. E$_{\rm{cm}}$ is the energy according to the relative velocity of the ion and the electron in the center of mass frame, and E$_{\rm{br}}$ is the released energy corresponding to the reaction channel. E$_v$ and E$_r$ are the vibrational and rotational energies, respectively. We assume all molecules and atoms to be in vibrational and rotational ground states. E$_{\rm{tot}}$ is then shared among the reaction products according to their masses. The components of the ion and electron velocity are chosen randomly from a 1D Maxwell-Boltzmann distribution according to the temperature of the ions and the electrons. Further details concerning the calculation of the energy of the produced hot atoms are given in \citet{groeller10}.

\subsection{Collisions between hot atoms and neutral background atmosphere}

 For the collisions between hot particles and the neutral background atmosphere, we use the same treatment as \citet{groeller14}. Basically, collisions can be elastic, inelastic, or quenching. While in elastic collisions the kinetic energy is conserved, in inelastic collisions, kinetic energy can be transformed into internal energy, i.e.\ vibrational energy or electronic excitation. During quenching collisions, an excited reactant will be de-excited and internal energy will be converted into kinetic energy.

Collisions between an energetic and a thermal O can be elastic or quenching. We do not consider inelastic collisions between two oxygen atoms since the energy released during the collision is smaller than the excitation energy. The total and differential cross sections for an elastic collision between a hot O atom in its ground state, O($^3$P), and an O atom of the background atmosphere are taken from \citet{tully01}. For elastic collisions between excited hot O atoms, O($^1$D) and O($^1$S), and a thermal O atom, the cross sections are taken from \citet{yee87} and \citet{yee85}, respectively. For the quenching O($^1$D)-O collisions, the total cross section is taken from \citet{yee90}, whereas the differential cross section is assumed to be the same as for the corresponding elastic collision. The cross sections for the quenching collision of a hot O($^1$S) with a thermal O are assumed to be the same as for the elastic collision.

Since there is no data for collisions between O and CO$_2$ or CO, we follow the approch of \citet{groeller14} and take the total and differential cross sections for O($^3$P,$^1$D,$^1$S)-N$_2$ collisions. For elastic O($^3$P)-N$_2$ collisions, the total cross section is taken from \citet{balakrishnan98b} and the differential cross section is the same as that for N-N$_2$ collisions, due to lack of other data \citep{balakrishnan98a}. For the corresponding inelastic collisions both cross sections are taken from \citet{balakrishnan98b}. The total cross section for the elastic collision of O($^1$D) and N$_2$ is taken from \citet{balakrishnan99}, whereas the differential cross section for this elastic collision and the cross sections for the corresponding inelastic collision are assumed to be the same as for O($^3$P)-N$_2$ collisions. For quenching O($^1$D)-N$_2$ collisions, the total cross section is from \citet{matsumi96} and the differential cross sections is taken to be the same as for O($^3$P). For all O($^1$S) collisions, the same cross sections as for O($^1$D) are employed.

We are not aware of cross sections for collisions between hot carbon atoms and thermal atoms or molecules of our background atmosphere. Hence, we employed the total and differential cross sections as given for hot oxygen atoms.

Since the simulated particles can reach energies of up to 10 eV, the total cross sections are extrapolated up to this value. In this study, we are only interested in the production and escape of hot atoms and not on their energy deposition into the background atmosphere. This means we do not consider any energy or momentum transfer to the background atmosphere. Therefore, any possible modification of the background atmosphere by the hot particles is neglected. However, a future study will focus on this process and its importance with changing EUV fluxes.

\section{Results} \label{sec:results}
\subsection{Loss rates of suprathermal atoms}

\begin{figure}
 \noindent\includegraphics[width=0.8\textwidth]{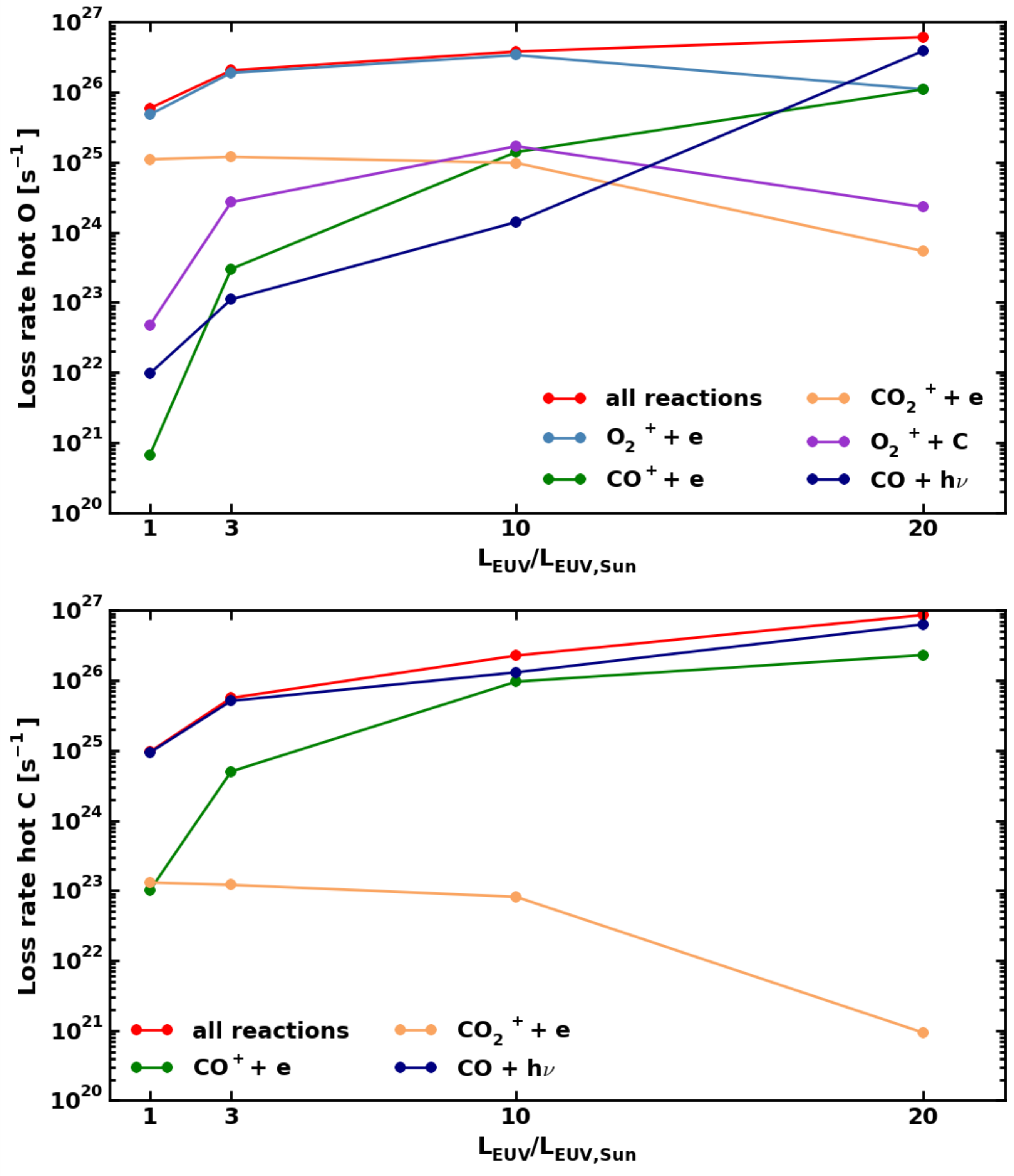}
 \caption{Loss rates of hot O (top) and hot C (bottom) as a function of EUV flux normalized to the present EUV flux of the Sun. The red curves display the total loss rates.}
 \label{fig:lossRates}
 \end{figure}

From the energy distribution functions at the exobase altitudes we determine the loss rates of hot oxygen and hot carbon. For the calculation of the loss rates we assume a uniform dayside escape flux and then integrate over $2\, \pi \, h_{exo}^2$, where $h_{exo}$ is the exobase altitude. The exobase altitudes, as calculated in our Monte-Carlo model, are about 220 km, 380 km, 750 km, and 5400 km for the 1, 3, 10, and 20 EUV case, respectively. Figure \ref{fig:lossRates} shows the loss rates of hot O (top) and hot C (bottom) as a function of the EUV flux. All reactions, apart from DR of CO$^+$ and PD of CO, decrease for a higher EUV flux than 10 times the present level. Such a behavior has also been reported by \citet{zhao15} for DR of O$_2^+$.

Basically, we have two competing mechanisms that seemingly have an influence on the importance of the production reaction for the loss of hot atoms. First, the production rates tend to increase with increasing EUV level, due to an increase in density of the involved species for the DR reactions and due to the rising EUV flux for the PD reactions. Second, due to the expansion of the atmosphere with higher EUV fluxes, there is a still significant expanded atmospheric layer above the main production zone, resulting in more collisions, and thus in increased energy loss of the hot particles on their way to the more distant exobase. We see in Figure \ref{fig:prod} that the production rates of PD of CO and DR of CO$^+$ for the 20 EUV case dominate above about 600 km, where the production rate of DR of CO$^+$ becomes larger than the one due to DR of O$_2^+$. Above this altitude, still a significant amount of primary hot particles is produced. The densities of the neutral background species, however, decrease with increasing altitude and correspondingly, collisions become lesser and lesser. Thus, most of these produced primary hot particles are able to reach the exobase at around 5400 km with energies larger than the escape energy. In these cases, the first mechanism (increase of production rate) dominates over the second one (expansion of atmosphere and thus more collisions). For all the other production reactions, the second mechanism is stronger than the first one, and the loss rates start to decrease when the EUV flux gets higher than 10 times the present level.

\begin{figure}
 \noindent\includegraphics[width=0.8\textwidth]{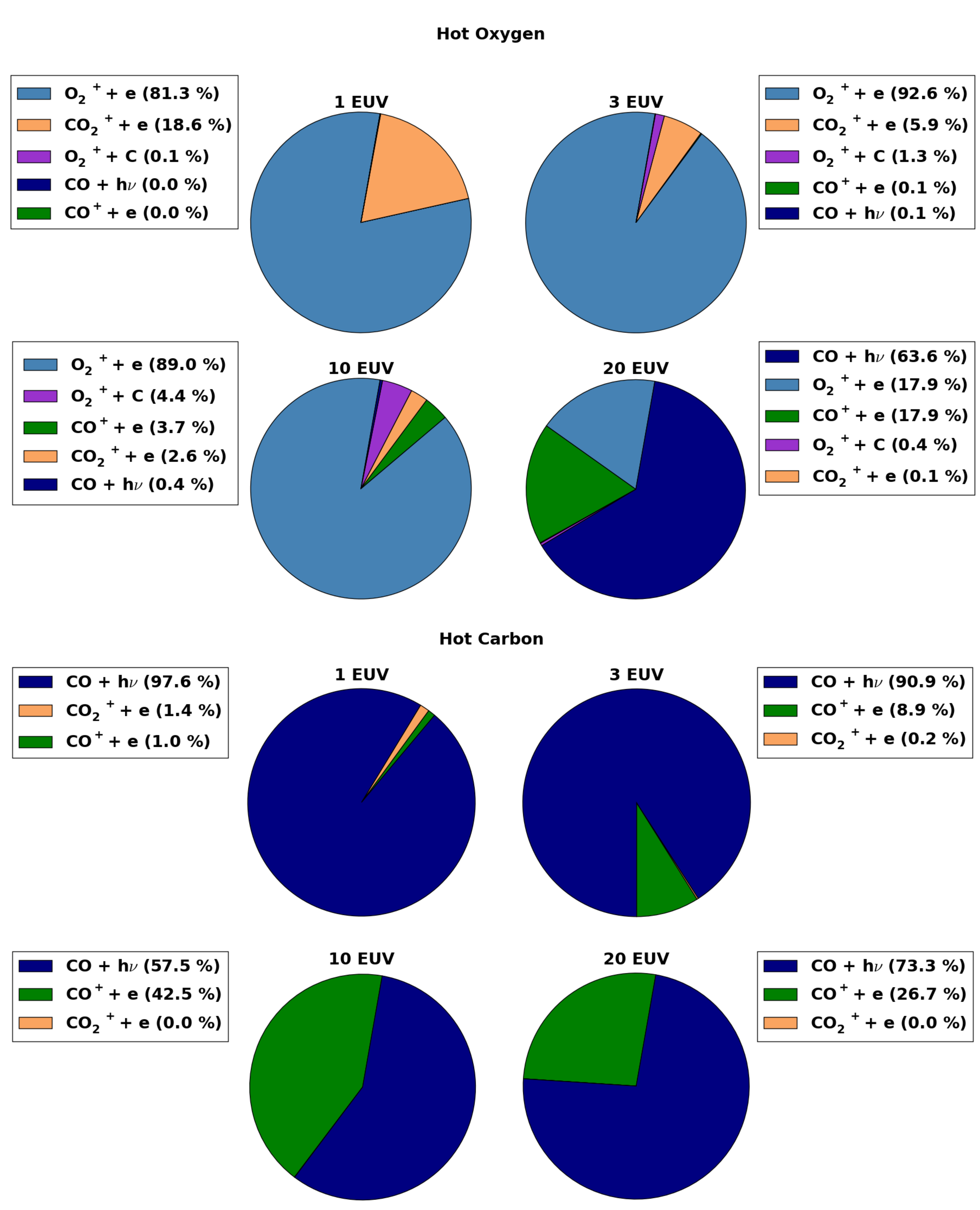}
 \caption{Contributions of the different production reactions of hot O (top panel) and hot C (bottom panel) to the loss for different EUV fluxes.}
 \label{fig:PieHot}
 \end{figure}

The importance of the different reactions regarding the loss rate for hot O and hot C is illustrated in Figure \ref{fig:PieHot}. While DR of O$_2^+$ is contributing most to the loss of hot O atoms in the 1, 3 and 10 EUV cases, PD of CO is dominating for the 20 EUV case. DR of CO$_2^+$ also contributes significantly (about $20$\%) to the escape of hot O atoms for the 1 EUV case, but becomes negligible for higher EUV cases. PD of CO is the major contributor to escaping hot C for all considered EUV cases, with DR of CO$^+$ becoming also important for the 10 and 20 EUV cases.

\begin{figure}
 \noindent\includegraphics[width=0.8\textwidth]{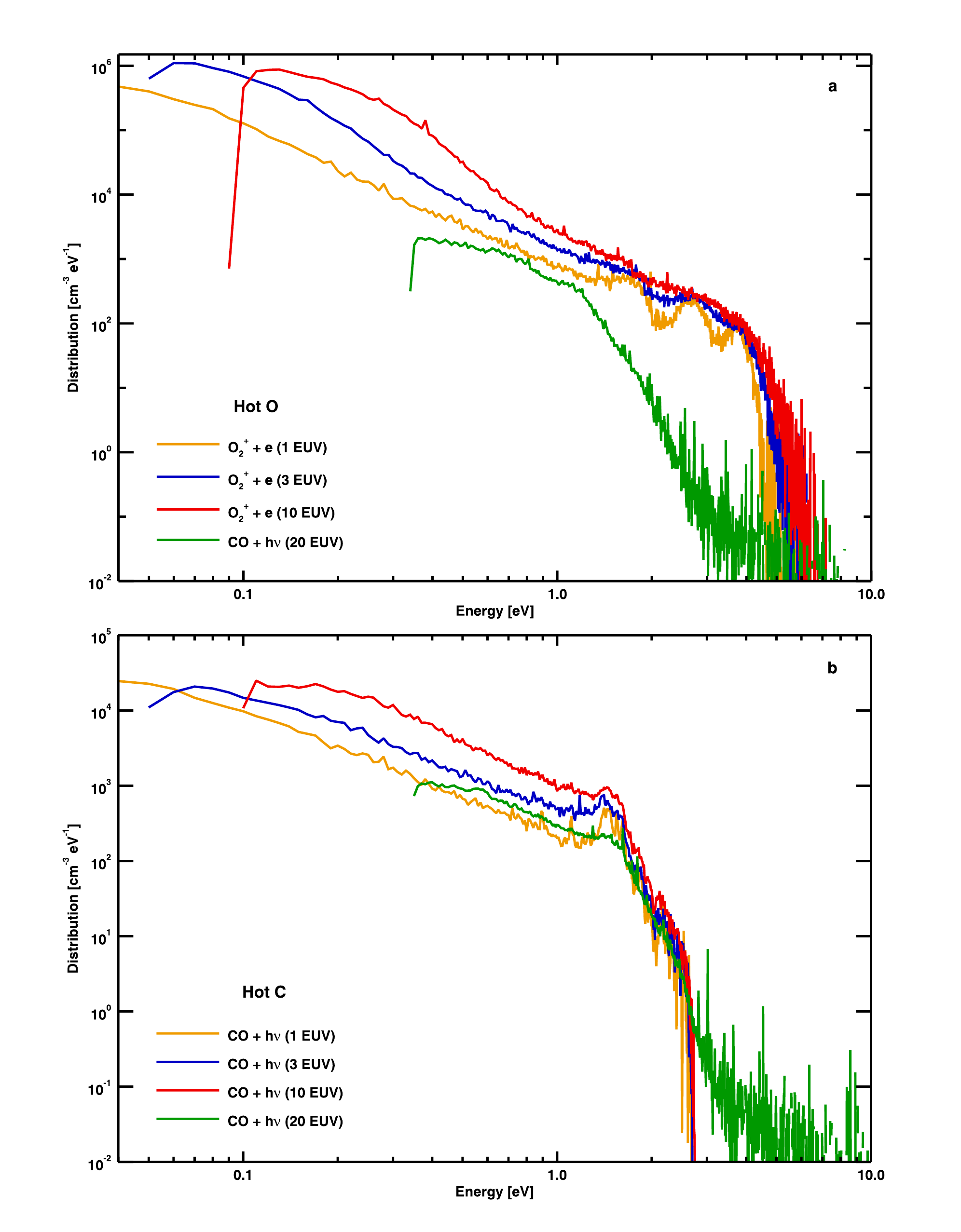}
 \caption{Energy distribution functions at the corresponding exobases for hot O (panel a) and hot C (panel b). Shown are the most important reactions for each of the EUV cases.}
 \label{fig:EDFs}
 \end{figure}

 Figure \ref{fig:EDFs} shows examples of the energy distribution functions (EDFs) at the corresponding exobases for hot O and hot C. Figure \ref{fig:EDFs}a shows the EDFs for hot O from DR of O$_2^+$, which is the most important reaction for 1, 3 and 10 EUV, and from PD of CO, which is the most important reaction in the 20 EUV case. Figure \ref{fig:EDFs}b displays the EDFs for hot C originating from PD of CO, which dominates all EUV cases. The cut-off energy at low energies is the energy corresponding to our stop condition for tracing hot particles, i.e.\ $1.5\,E_\mathrm{therm}$, which of course is different for the different EUV cases. The cut-off energies are all well below the corresponding escape energies, which are about $2$ eV for the 1 and 3 EUV cases, about $1.7$ eV for the 10 EUV case, and about $0.7$ eV for the 20 EUV case.

Table \ref{tab:lossRates} summarizes the loss rates for hot oxygen and hot carbon for the reactions considered in this study. Especially the loss rate of hot C due to PD of CO in the 1 EUV case is higher than the loss rates of previous studies, e.g. \citet{fox01}, \citet{nagy01}, and \citet{groeller14}. \citet{fox01} used a completely different approch, the ``exobase approximation'', in which only those hot atoms produced above the exobase contribute to the loss. As mentioned in their study, this approach underestimates the suprathermal loss due to PD of CO, since the photodissociation rate decreases with increasing altitude, and thus we can assume that a large portion of escaping hot C may come from altitudes below the exobase. Additionally, they used different photoabsorption cross sections and a different solar flux than we do in our study. \citet{nagy01} obtained a higher escape flux than \citet{fox01}, but still a lower one than ours. They also used a different method, the ``two-stream model'', to obtain their escape fluxes, and a different solar flux as well as a different cross section for photodissociation than we do in our study. It should also be noted that in our Monte-Carlo simulations newly produced secondary hot C atoms also contribute to the loss rate of PD of CO, whereas in \citet{fox01} the escape rates/fluxes are due to only the reaction itself and not due to secondary produced hot C atoms. In the present study, we use the same Monte-Carlo model and collision cross sections described in \citet{groeller14}, only the solar flux and the input profiles are different. The ion profiles used in our study compare to the low solar activity eroded profiles of \citet{fox09}, used by \citet{groeller14}. The neutral profiles of our study lie in between the low solar activity and high solar activity profiles of \citet{fox09}.  The use of a different solar flux and different input profiles results in higher loss rates than in \citet{groeller14}. The main difference comes from the use of the SUMER quiet Sun solar flux. We take these measurements, because they have a very high resolution. Also, since we do not know the activity of the Sun in earlier times, we decided to take the high resolved flux of the quiet Sun conditions for today and for higher EUV fluxes (i.e. earlier times).

\subsection{Loss of atmospheric CO$_2$ pressure through suprathermal atoms}

To get the total loss of hot O and hot C and the corresponding amount of atmospheric pressure over approximately the last 4 Gyr, we integrate the interpolated loss rates over time. Interpolation is done linearly between the times corresponding to the considered EUV fluxes. For the loss of CO$_2$, the loss of C is an important factor, since C (suprathermal or not) is definitely produced primarily from CO$_2$, whereas O can also originate from dissociation of H$_2$O. Thus, the loss of C is a direct indication of the loss of CO$_2$. For the calculation of the lost CO$_2$, we assume that for one escaping C atom, we have two escaping O atoms. Table \ref{tab:atmPress} shows the loss of atmosphere pressure, i.e.\ loss of CO$_2$, from different times in the past until today for the three different rotator cases.

From Table \ref{tab:lossRates} we see that the 2:1 relation for O:C is not fulfilled for the 10 and 20 EUV cases. For the 20 EUV case the loss of suprathermal C is even larger than the loss of suprathermal O. Thus, oxygen additionally has to be lost through other processes, as there are for example ion pick-up \citep[e.g.][]{curry13} or chemical surface weathering, such as oxidation \citep[e.g.][]{lammer03a,lammer13}. This means that the lost CO$_2$ pressures for the 10 and 20 EUV cases as calculated in this study have to be seen as maximum values that can be lost through suprathermal atoms.

\section{Surface pressure evolution of Mars' CO$_2$ atmosphere}\label{sec:surfPress}

\begin{figure}
 \noindent\includegraphics[width=0.8\textwidth]{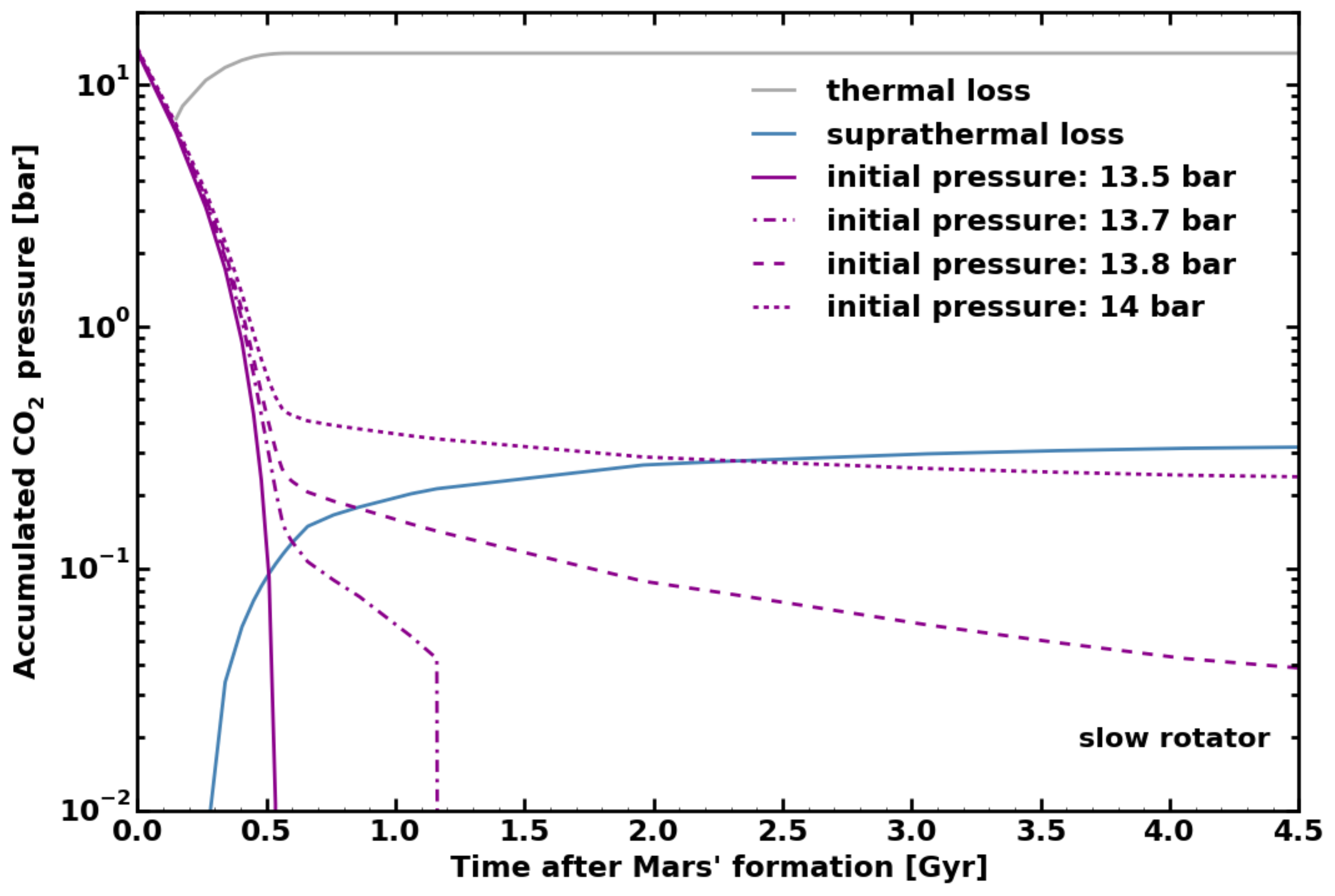}
 \caption{Evolution of the martian surface pressure as a result of an initially magma ocean related outgassed CO$_2$ atmosphere between 13 and 14 bar and thermal as well as suprathermal loss. The young Sun was assumed to be a slow rotator.}
 \label{fig:EvolInit}
 \end{figure}

 Catastrophic outgassing of an initial steam atmosphere, related to a magma ocean, may have occured after proto-Mars finished its formation within the first 10 Myr after the origin of the Sun \citep[e.g.][]{walsh11,brasser13}. Consequently, condensation of H$_2$O and the possible formation of large lakes or even an ocean in a warm and wet environment could have happened \citep{hamano13,lebrun13}. Figure \ref{fig:EvolInit} shows the surface pressure evolution (magenta curves) resulting from thermal (grey curve) and suprathermal (blue curve) CO$_2$ escape and assuming different magma ocean related outgassed CO$_2$ amounts between 13 and 14 bar (we discuss in the next paragraph, why these specific values are taken). The radiation evolution of the Sun corresponds to a slow rotator for this figure. The thermal loss rates are taken from the study of \citet{tian09}. For this figure, we take an initial CO$_2$ amount (13.5, 13.7, 13.8, and 14 bar) and subtract the lost CO$_2$ pressure due to thermal and suprathermal loss processes. In the beginning of Mars' atmospheric evolution, thermal loss is much higher than suprathermal loss. The initially outgassed CO$_2$ atmosphere is significantly reduced due to thermal loss processes. However, the magma ocean related outgassed CO$_2$ atmosphere would have been larger than 1 bar during the pre-Noachian epoch (until about 0.4 Gyr after Mars' formation). Thus, Mars may have had standing bodies of liquid water during the pre-Noachian era due to the post-magma ocean surface temperatures and the possibility of the H$_2$-CO$_2$ greenhouse effect \citep{ramirez14}.

According to the results of the study by \citet{tian09}, which is based on the atmospheric background also used in the present study, C atoms flow out hydrodynamically with a loss rate of the order of about $10^{30}$ s$^{-1}$ from the EUV heated and extended exobase level for a 20 times higher EUV flux. A slight decrease of the EUV flux enhancement reduces the thermal escape rates dramatically. For the 10 EUV case, the C and O loss rates are in the Jeans escape domain and negligible compared to the suprathermal loss rates of about $2.3\times 10^{26}$ s$^{-1}$ and $3.8\times 10^{26}$ s$^{-1}$, respectively, shown in Table \ref{tab:lossRates}. Thus, in the late Noachian era (about $0.6 - 0.8$ Gyr after Mars' formation), the loss of suprathermal atoms takes over (Figure \ref{fig:EvolInit}). An initially 14 bar CO$_2$ atmosphere, which is lessened due to thermal loss to about 500 mbar at 0.6 Gyr after Mars formed, would not have completely escaped to space until today via suprathermal loss processes and approximately 220 mbar CO$_2$ would have remained (Figure \ref{fig:EvolInit}). Thus, our results suggest that 14 bar are an upper limit for an initially outgassed CO$_2$ atmosphere, when considering thermal and suprathermal loss. Initial atmospheres with pressure values of less than 13.7 bar could have been lost completely by a combination of thermal and suprathermal loss. For similar initial outgassing amounts but higher EUV fluxes, as would be the case for a moderate or fast rotating young Sun, the initial martian CO$_2$ atmosphere would have vanished much earlier resulting in a dry and cold environment during most of the pre-Noachian era.

To estimate different volcanic outgassing scenarios, we use a mantle convection model for a 2D spherical annulus \citep{noack16} to investigate the thermal evolution of Mars and consequent volcanic outgassing of CO$_2$. The model parameters mainly follow \citet{grott11}, where we assume that 100 ppm of CO$_2$ in the melt (limited by oxygen fugacity of Mars' mantle) and 40\% extrusive volcanism.

\begin{figure}
 \noindent\includegraphics[width=0.8\textwidth]{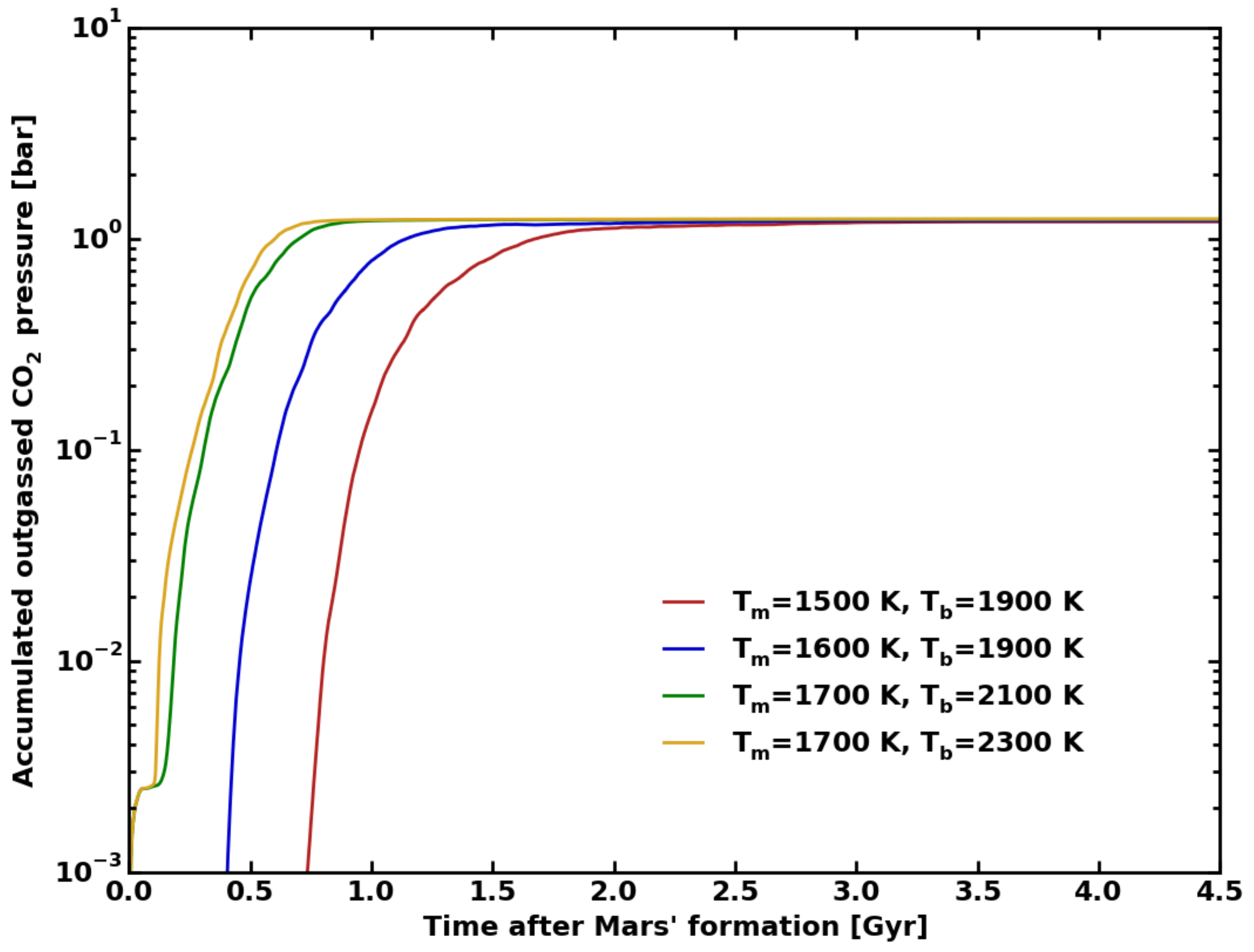}
 \caption{Different volcanic CO$_2$ outgassing scenarios. The different cases refer to different assumptions of the initial bottom temperature at the core-mantle boundary for different superheated core scenarios ($T_{\rm b}$) and the initial mantle temperature below an initially 100 km thick lithosphere ($T_{\rm m}$).}
 \label{fig:OutgCO2}
 \end{figure}

\begin{figure}
 \noindent\includegraphics[width=0.8\textwidth]{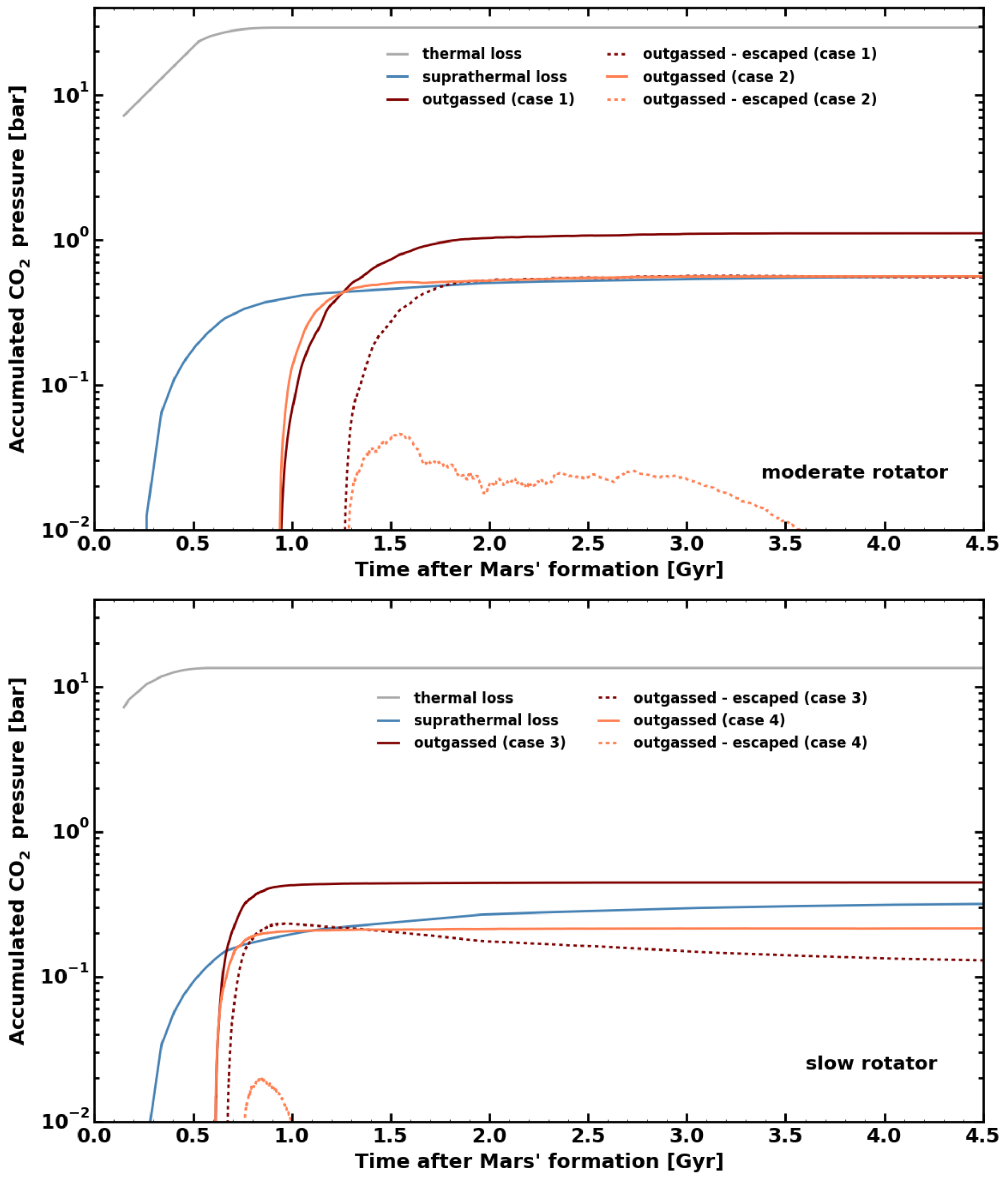}
 \caption{Evolution of the martian surface pressure as a result of volcanic outgassing and thermal as well as suprathermal loss for a moderate rotating (top) and slow rotating (bottom) young Sun. The different cases are described in the text.}
 \label{fig:EvolOutg}
 \end{figure}

Figure \ref{fig:OutgCO2} shows a few scenarios of volcanic CO$_2$ outgassing with different initial bottom temperatures at the core-mantle boundary for different superheated core scenarios ($T_{\rm b}$) \citep{breuer03} and initial mantle temperatures below an initially 100 km thick lithosphere ($T_{\rm m}$). We consider four cases:
\begin{itemize}
 \item Case 1: $T_{\rm m} = 1500$ K, $T_{\rm b} = 1900$ K
 \item Case 2: $T_{\rm m} = 1600$ K, $T_{\rm b} = 1900$ K
 \item Case 3: $T_{\rm m} = 1700$ K, $T_{\rm b} = 2100$ K
 \item Case 4: $T_{\rm m} = 1700$ K, $T_{\rm b} = 2300$ K
\end{itemize}
We see that the larger $T_{\rm b}$, the earlier the outgassing starts. Figure \ref{fig:EvolOutg} represents the CO$_2$ presssure evolution due to different volcanic outgassing cases and escape to space (top: moderate rotator; bottom: slow rotator; the evolution of the fast and moderate rotator do not differ much for the EUV fluxes considered here, thus we only compare two cases). Due to the high thermal loss rates (gray curves), the outgassed CO$_2$ amount before 0.6 (slow rotator) and 0.94 (moderate rotator) Gyr after Mars' formation cannot accumulate in the atmosphere, but is lost to space. Thus, the accumulation of outgassed CO$_2$ pressure starts when the thermal loss has ceased. Everything that is outgassed from then on, will be reduced due to loss through suprathermal atoms (blue curves). If too much gets outgassed at the end of the Noachian or beginning of the Hesperian era (case 1 for moderate rotator, case 3 for slow rotator), it cannot be lost through suprathermal atoms. If too less gets outgassed (case 2 for moderate rotator, case 4 for slow rotator), then all of the outgassed CO$_2$ can be lost by suprathermal loss. If about 250 mbar of atmospheric CO$_2$ pressure were in the martian atmosphere at around 1 Gyr after Mars formed for a slow rotator (case 3), then there would be approximately 100 mbar in today's atmosphere --- if there had been escape only through suprathermal atoms, which was certainly not the case. For a moderate rotator, the atmospheric pressure at about 1.5 Gyr after Mars' formation can be larger, since more is lost due to higher loss rates at these times, corresponding to different EUV flux levels. Our results show a higher atmospheric pressure in the past than the value given by \citet{gillmann11}, who predict about 50 mbar about 4 Gyr ago. One reason for this difference is that our loss rate due to suprathermal atoms is higher than their loss rate, which was extrapolated to the past assuming the EUV flux evolution of \citet{ribas05}. In agreement with \citet{gillmann11} our study also indicates that most of a primordial atmosphere must have been lost thermally rather rapidly within the first 0.5 to 1 Gyr after Mars' formation, since it could not have been removed by non-thermal loss processes afterwards, and that today's atmosphere of Mars is built up mostly of volcanic gases, outgassed after the loss of the primordial atmosphere.

The time span between the decrease of the EUV flux from 20 to 10 EUV is an important factor. If one assumes that the young Sun was a fast rotator, the 20 EUV case would have been reached at about 0.8 Gyr after Mars' formation with a decrease of the EUV flux to about 10 EUV at about 1.1 Gyr. However, this fast rotator case is very unlikely. If the solar EUV flux was 20 EUV at 0.8 Gyr and 10 EUV at 1.1 Gyr after Mars formed, Earth would have lost its nitrogen atmosphere during the Archean epoch, because of the absence of sufficient amounts of CO$_2$ required as a thermospheric IR-cooler \citep{lichtenegger10}. The mineralogy of Archaean sediments, such as the ubiquitous presence of mixed-valence Fe(II-III) oxides (magnetite) in banded iron formations \citep[e.g.][]{rosing10} is also inconsistent with the necessary amount of CO$_2$ that would act as a cooler of the upper atmosphere against the high EUV fluxes related to a fast rotator during that time period. The same is true for the moderate rotator case, whose radition evolution does not differ much from a fast rotator.

 Nonetheless, even if we assume that the early Sun was a slow rotator and Mars experienced a hotter and wetter period during the pre-Noachian era, after the main loss of its initial CO$_2$ inventory, loss due to hot atoms most likely eroded not more than about 150 mbar during the past 3.7 Gyr. In a follow-up study, we will investigate the solar wind induced escape of planetary C and O from the hot atom coronae in the past (ion pick-up and sputtering). These processes could additionally increase the loss of CO$_2$ over the last 4 Gyr, which means that the surface pressure could have been higher at the end of the Noachian era. Additionally to the escape of CO$_2$ to space, a few 10 to 100 mbar may have also been lost to the surface during the Hesperian and Amazonian epochs due to carbonate precipitation and serpentinization \citep[e.g.][]{chassefiere11}. If we roughly estimate that ion pick-up and sputtering are of the same order of magnitude and surface weathering less effective than suprathermal loss processes, the amount of CO$_2$ present in the martian atmosphere at the end of the Noachian epoch was probably not more than about 500 mbar.

\section{Summary and conclusions} \label{sec:summary}

We conducted Monte-Carlo simulations of hot oxygen and hot carbon in the atmosphere of Mars for different EUV fluxes (1, 3, 10, and 20 EUV), corresponding to different times in the past. We use a background atmosphere with four neutral species (O, CO, CO$_2$, C) and four ion species (O$_2^+$, CO$_2^+$, CO$^+$, O$^+$). For the production of the hot atoms we consider five source reactions for oxygen and three reactions for carbon. We calculate the energy distribution functions at the exobase altitude and determine the loss rates of hot oxygen and hot carbon. For hot oxygen, DR of O$_2^+$ appears to be always an important contribution to the loss of hot O, while DR of CO$^+$ and PD of CO become significant only in the 20 EUV case. For hot carbon, PD of CO is the major contributor to escaping hot C for all EUV cases, with DR of CO$^+$ becoming also important for the 10 and 20 EUV levels. The loss rates due to DR of CO$^+$ and PD of CO increase with increasing EUV level, whereas the escape due to the other reactions decreases when the EUV flux is higher than 10 times the present level.

Considering different possible radiation evolution models for the Sun, i.e.~the slow, moderate, and fast rotator, we can relate the different EUV levels to different times in the past of the martian history. Taking these points in time, we integrated the loss rates and estimated the lost atmospheric CO$_2$ pressure due to hot atoms. For this estimation we assumed a 2:1 relation for lost hot O to lost hot C. Depending on the radiation evolution of the Sun, we find that atmospheric pressures ranging from approximately 200 to 400 mbar were able to escape to space during the last 4 Gyr (Table \ref{tab:atmPress}). There are good reasons to believe that the Earth could not have kept its nitrogen atmosphere if the early Sun was a moderate or fast rotator, as discussed in Section \ref{sec:surfPress}. Therefore, assuming that the Sun has been a slow rotator, our results indicate that Mars could not have had a significant CO$_2$ atmosphere at the end of the Noachian epoch. If the atmosphere was denser, it could not have been lost through non-thermal loss and surface weathering processes until today. In the pre-Noachian era, however, Mars could have had a magma ocean related outgassed CO$_2$ atmosphere of a few bar for about 300 Myr.


%

\begin{acknowledgments}
U.V. Amerstorfer and H. Lichtenegger are supported by the Austrian Science Fund (FWF): P24247-N16. H. Lammer, M. G\"udel and C. Johnstone acknowledge support by the Austrian Science Fund (FWF): S11601-N16, S11604-N16, and S11607-N16. M. Scherf is supported by the Austrian Science Fund (FWF): S11606-N16. L. Noack has been funded by the Interuniversity Attraction Poles Programme initiated by the Belgian Science Policy Office through the Planet Topers alliance. Finally, C. Johnstone, M. G\"udel, H. Lammer, and L. Noack thank the International Space Science Institute (ISSI) in Bern, and the ISSI team ``The Early Evolution of the Atmospheres of Earth, Venus, and Mars''. The data for the results of this study can be downloaded under http://www.iwf.oeaw.ac.at/JGR-2016JE005175R/ and are available from the authors upon request (ute.amerstorfer@oeaw.ac.at)
\end{acknowledgments}

%
%
\end{article}
\newpage

 \newpage

 \newpage

 \newpage

 \newpage

 \newpage

 \newpage

 \newpage

 \newpage

 \newpage

 \newpage

 \begin{table}
 \caption{Times in the past for different EUV levels and different rotation evolution of the young Sun. The Noachian epoch lasts from 4.1 to 3.7 Gyr ago, and the Hesperian from 3.7 to 2.0 Gyr ago.}
 \label{tab:rotEUVTimes}
 \centering
  \begin{tabular}{c c c c c}
  \hline \\[-0.3cm]
   & 3  & 10  & 20 & EUV \\[0.1cm]
\hline \\[-0.3cm]
Slow rotator & 2.6 & 3.8 & 4.3 & [Gyr ago]\\[0.2cm]
Moderate rotator & 2.3 & 3.5 & 3.9  & [Gyr ago]\\[0.2cm]
Fast rotator & 2.3 & 3.4 & 3.7 & [Gyr ago]\\[0.2cm]
  \hline
  \end{tabular}
 \end{table}

\begin{table}
\caption{Sources of hot O and C, and their coefficients for the production rate.}
\label{tab:sources}
\centering
 \begin{tabular}{l l}
  \hline \\[-0.4cm]
  Source reaction hot O & Coefficients \\[0.1cm]
  \hline \\[-0.3cm]
  Dissociative recombination &  \\[0.2cm]
  O$_2^+$ + e $\to$ O + O &  $T_{\rm e} \leq 1200$ K: $\alpha = 2.4 \times 10^{-7}$ cm$^{3}$ s$^{-1}$, $\beta = -0.7$\\[0.2cm]
  						  &  $T_{\rm e} > 1200$ K: $\alpha = 1.9 \times 10^{-7}$ cm$^{3}$ s$^{-1}$, $\beta = -0.61$\\[0.2cm]
						  &  \citep{sheehan04}\\[0.2cm]
  CO$^+$ + e $\to$ C + O & $\alpha = 2.75 \times 10^{-7}$ cm$^{3}$ s$^{-1}$, $\beta = -0.55$ \\[0.2cm]
					     & \citep{rosen98}\\[0.2cm]
  CO$_2^+$ + e $\to$ CO + O & $\alpha = 4.2 \times 10^{-7}$ cm$^{3}$ s$^{-1}$, $\beta = -0.75$ \\[0.2cm]
  							& branching ratio: 96\% \\[0.2cm]
					        & \citep{viggiano05}\\[0.2cm]
  Chemical reaction &  \\[0.2cm]
  O$_2^+$ + C $\to$ CO$^+$ + O & $k = 5.2 \times 10^{-11}$ cm$^{3}$ s$^{-1}$, \\[0.2cm]
					           & \citep{mcelroy13}\\[0.2cm]
  \hline
  \hline \\[-0.4cm]
  Source reaction hot C & Coefficients \\[0.1cm]
  \hline \\[-0.3cm]
  Dissociative recombination &  \\[0.2cm]
  CO$^+$ + e $\to$ O + C & $\alpha = 2.75 \times 10^{-7}$ cm$^{3}$ s$^{-1}$, $\beta = -0.55$ \\[0.2cm]
					       & \citep{rosen98}\\[0.2cm]
  CO$_2^+$ + e $\to$ O$_2$ + C  & $\alpha = 4.2 \times 10^{-7}$ cm$^{3}$ s$^{-1}$, $\beta = -0.75$ \\[0.2cm]
  								& branching ratio: 4\% \\[0.2cm]
					            & \citep{viggiano05}\\[0.2cm]
  \hline
 \end{tabular}
\end{table}
\newpage

\begin{table}
\caption{Loss rates of hot oxygen and hot carbon atoms.}
\label{tab:lossRates}
\centering
\begin{tabular}{l c c c c}
\hline \\[-0.4cm]
 & \multicolumn{3}{c}{ Loss rate [s$^{-1}$]} \\
Hot O & 1 EUV & 3 EUV & 10 EUV & 20 EUV \\[0.1cm]
\hline \\[-0.3cm]
 CO + $h\nu$ $\to$ C + O & $9.6 \times 10^{21}$ & $1.1 \times 10^{23}$ & $1.4 \times 10^{24}$ & $3.9 \times 10^{26}$\\[0.2cm]
 O$_2^+$ + e $\to$ O + O & $4.8 \times 10^{25}$ & $1.9 \times 10^{26}$ & $3.4 \times 10^{26}$ & $1.1 \times 10^{26}$\\[0.2cm]
 CO$^+$ + e $\to$ C + O & $6.6 \times 10^{20}$ & $3.0 \times 10^{23}$ & $1.4 \times 10^{25}$ & $1.1 \times 10^{26}$\\[0.2cm]
 CO$_2^+$ + e $\to$ CO + O & $1.1 \times 10^{25}$ & $1.2 \times 10^{25}$ & $9.8 \times 10^{24}$ & $5.4 \times 10^{23}$\\[0.2cm]
 O$_2^+$ + C $\to$ CO$^+$ + O & $4.7 \times 10^{22}$ & $2.7 \times 10^{24}$ & $1.7 \times 10^{25}$ & $2.3 \times 10^{24}$\\[0.2cm]
 \hline \\[-0.3cm]
 Total & $5.9 \times 10^{25}$ & $2.1 \times 10^{26}$ & $3.8 \times 10^{26}$ & $6.1 \times 10^{26}$\\[0.2cm]
 \hline \\[-0.3cm]
Hot C &  &  &  \\[0.1cm]
\hline \\[-0.3cm]
 CO + $h\nu$ $\to$ O + C & $9.3 \times 10^{24}$ & $5.1 \times 10^{25}$ & $1.3 \times 10^{26}$ & $6.3 \times 10^{26}$\\[0.2cm]
 CO$^+$ + e $\to$ O + C & $1.0 \times 10^{23}$ & $5.0 \times 10^{24}$ & $9.6 \times 10^{25}$ & $2.3 \times 10^{26}$\\[0.2cm]
 CO$_2^+$ + e $\to$ O$_2$ + C & $1.3 \times 10^{23}$ & $1.2 \times 10^{23}$ & $8.1 \times 10^{22}$ & $9.3 \times 10^{20}$\\[0.2cm]
 \hline \\[-0.3cm]
 Total & $9.5 \times 10^{24}$ & $5.6 \times 10^{25}$ & $2.3 \times 10^{26}$ & $8.6 \times 10^{26}$\\ [0.2cm]
 \hline
\end{tabular}
\end{table}
\newpage

 \begin{table}
 \caption{Loss of atmospheric CO$_2$ pressure due to hot atoms for different times in the past until today for a slow, moderate and fast rotator. The slanted values for the moderate and fast rotator cases beyond 3.9 and 3.7 Gyr ago, respectively, are for EUV fluxes larger than 20 times the present solar flux. These values were linearly extrapolated, since we do not have input data for simulations. As discussed in the text, extrapolation of loss rates into past times is a rather insecure method, and hence, these extrapolated values have to be taken with care. The Noachian era is from 4.1 to 3.7 Gyr ago, and the Hesperian era from 3.7 to 2.0 Gyr ago.}
 \label{tab:atmPress}
 \centering
  \begin{tabular}{c c c c c c c c c c}
  \hline \\[-0.3cm]
   & 2.3  & 2.6  & 3.4  & 3.5 & 3.7 & 3.8  & 3.9  & 4.3 & [Gyr ago]  \\[0.1cm]
\hline \\[-0.3cm]
Slow rotator & 41 & 51 & 104  & 115  & 138  & 151  & 168  & 312 &[mbar]\\[0.2cm]
Moderate rotator & 45 & 59 & 132  & 145  & 191  & 228  & 274  & {\it 553} &[mbar]\\[0.2cm]
Fast rotator & 45 & 59 & 137  & 157  & 234  & {\it 291}  & {\it 361}  & {\it 765} &[mbar]\\[0.2cm]
  \hline
  \end{tabular}
 \end{table}

\end{document}